\documentclass[trackchanges,twocolumn]{aastex701}

\usepackage{graphicx}
\usepackage{amsmath}

\begin{document}

\title{High-resolution Observations of Physical Properties of Dynamic Fibrils Around a Sunspot}

\author{Guotang Wu}
\affiliation{Yunnan Observatories, Chinese Academy of Sciences, Kunming 650216, People's Republic of China}
\affiliation{University of Chinese Academy of Sciences, Beijing 100049, People's Republic of China}
\email{wuguotang@ynao.ac.cn}

\author[0000-0003-2891-6267]{Xiaoli Yan}
\affiliation{Yunnan Observatories, Chinese Academy of Sciences, Kunming 650216, People's Republic of China}
\affiliation{Yunnan Key Laboratory of Solar Physics and Space Science, Kunming 650216, People's Republic of China}
\email{yanxl@ynao.ac.cn}

\correspondingauthor{Xiaoli Yan}
\email{yanxl@ynao.ac.cn}

\author[0000-0002-6526-5363]{Zhike Xue}
\affiliation{Yunnan Observatories, Chinese Academy of Sciences, Kunming 650216, People's Republic of China}
\affiliation{Yunnan Key Laboratory of Solar Physics and Space Science, Kunming 650216, People's Republic of China}
\email{zkxue@ynao.ac.cn}

\author[0000-0003-4393-9731]{Jincheng Wang}
\affiliation{Yunnan Observatories, Chinese Academy of Sciences, Kunming 650216, People's Republic of China}
\affiliation{Yunnan Key Laboratory of Solar Physics and Space Science, Kunming 650216, People's Republic of China}
\email{wangjincheng@ynao.ac.cn}

\author[0000-0002-9121-9686]{Zhe Xu}
\affiliation{Yunnan Observatories, Chinese Academy of Sciences, Kunming 650216, People's Republic of China}
\affiliation{Yunnan Key Laboratory of Solar Physics and Space Science, Kunming 650216, People's Republic of China}
\email{xuzhe6249@ynao.ac.cn}

\author[0000-0003-0236-2243]{Liheng Yang}
\affiliation{Yunnan Observatories, Chinese Academy of Sciences, Kunming 650216, People's Republic of China}
\affiliation{Yunnan Key Laboratory of Solar Physics and Space Science, Kunming 650216, People's Republic of China}
\email{yangliheng@ynao.ac.cn}

\author[0000-0001-9491-699X]{Yadan Duan}
\affiliation{Yunnan Observatories, Chinese Academy of Sciences, Kunming 650216, People's Republic of China}
\affiliation{Yunnan Key Laboratory of Solar Physics and Space Science, Kunming 650216, People's Republic of China}
\email{duanyadan@ynao.ac.cn}

\author[0000-0002-0464-6760]{Yian Zhou}
\affiliation{Yunnan Observatories, Chinese Academy of Sciences, Kunming 650216, People's Republic of China}
\affiliation{Yunnan Key Laboratory of Solar Physics and Space Science, Kunming 650216, People's Republic of China}
\email{zhouyian@ynao.ac.cn}

\author{Xinsheng Zhang}
\affiliation{Yunnan Observatories, Chinese Academy of Sciences, Kunming 650216, People's Republic of China}
\affiliation{University of Chinese Academy of Sciences, Beijing 100049, People's Republic of China}
\email{zhangxinsheng@ynao.ac.cn}

\author{Qifan Dong}
\affiliation{Yunnan Observatories, Chinese Academy of Sciences, Kunming 650216, People's Republic of China}
\affiliation{University of Chinese Academy of Sciences, Beijing 100049, People's Republic of China}
\email{dongqifan@ynao.ac.cn}

\author[0009-0003-9377-1989]{Zongyin Wu}
\affiliation{Yunnan Observatories, Chinese Academy of Sciences, Kunming 650216, People's Republic of China}
\affiliation{University of Chinese Academy of Sciences, Beijing 100049, People's Republic of China}
\email{wuzongyin@ynao.ac.cn}

\begin{abstract}

Understanding the formation and evolution of dynamic fibrils (DFs) in active regions is essential for uncovering the physical processes operating in the highly structured and magnetized solar atmosphere. High-resolution chromospheric observations from the New Vacuum Solar Telescope (NVST) show DFs with periodic up-and-down motions. We carry out a statistical analysis of 176 DFs around a sunspot. The mean values of deceleration, maximum velocity, maximum length, and duration of the DFs are $120 \pm 43~\mathrm{m~s^{-2}}$, $15.9 \pm 4.5~\mathrm{km~s^{-1}}$, $1.13 \pm 0.48~\mathrm{Mm}$, and $270 \pm 63~\mathrm{s}$, respectively. Several pairs of parameters exhibit correlations, most notably a clear positive correlation between deceleration and maximum velocity. The consistency between the periodicities of DFs and running penumbral waves is first identified. We further find that DFs in regions with more vertical magnetic fields exhibit larger decelerations and shorter periods, whereas those in regions with more horizontal magnetic fields exhibit the opposite behavior. These results, together with the presence of running penumbral waves, support the interpretation that DFs are driven by slow magnetoacoustic waves. Furthermore, the multiwavelength signatures of DFs from the chromosphere to the corona imply a potential contribution to energy transport and heating in the solar atmosphere.

\end{abstract}

\keywords{
\href{https://astrothesaurus.org/uat/1475}{Solar activity (1475)};
\href{https://astrothesaurus.org/uat/1479}{Solar chromosphere (1479)};
\href{https://astrothesaurus.org/uat/1515}{Solar oscillations (1515)}
}

\section{Introduction} \label{sec:section1} 

Dynamic fibrils (DFs) are elongated, jet-like chromospheric features observed in solar active regions, characterized by their periodic up-and-down motions (see reviews by \citealp{2007ASPC..368...27R} and \citealp{2019ARA&A..57..189C}). In recent years, a series of detailed studies of DFs has emerged from advances in high-resolution observations and numerical simulations (e.g., \citealp{2006ApJ...647L..73H,2006A&A...460..309D,2007ApJ...655..624D,2008ApJ...673.1194L,2008ApJ...673.1201L,2011ApJ...739...92P,2013ApJ...776...56R}). Extensive evidence suggests that DFs are the active region counterparts of spicules observed at the solar limb and mottles observed on the solar disk, with similar morphologies and characteristics. Such small-scale dynamic phenomena are therefore crucial to understanding the dynamics of the solar atmosphere.

\begin{figure*}[ht]
    \centering
    \includegraphics[width=\linewidth]{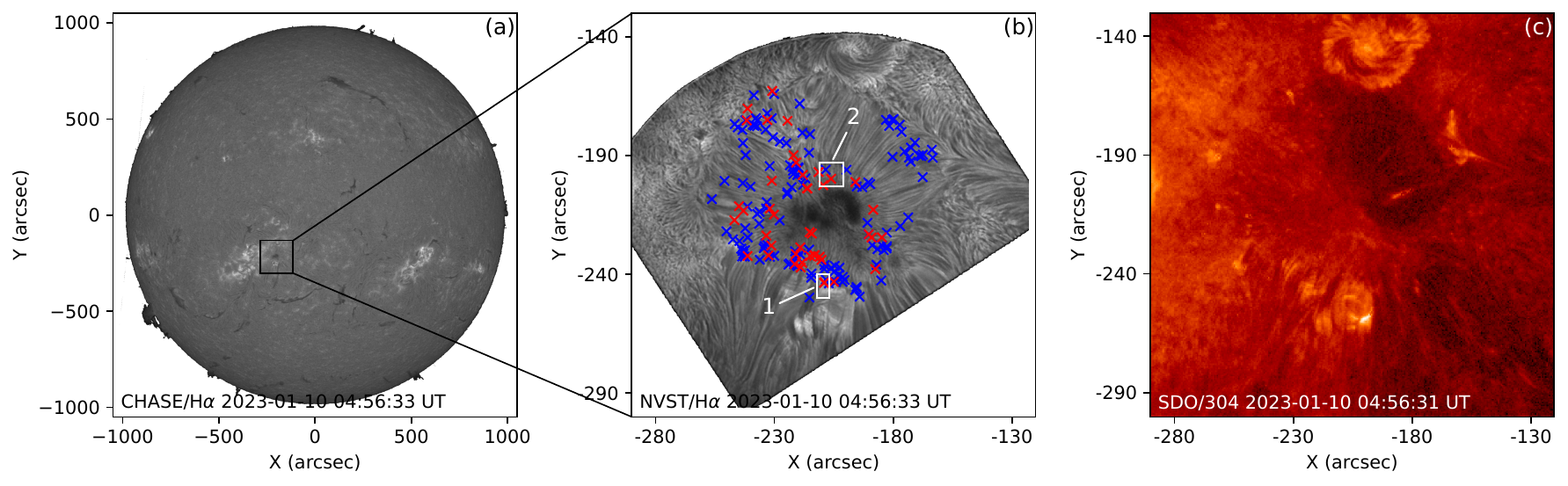}
    \caption{Overview of active region NOAA 13182. (a) Full-disk H$\alpha$ image from CHASE. The black box marks the field of view of NVST. (b) H$\alpha$ image from NVST. Blue crosses denote single-parabola events, while red crosses denote multiple-parabola events. The white boxes mark subregions used in subsequent figures. (c) Corresponding 304\,\AA\ image from SDO.}
    \label{fig:figure1}
\end{figure*}

\begin{figure*}[ht]
    \centering
    \includegraphics[width=0.78\linewidth]{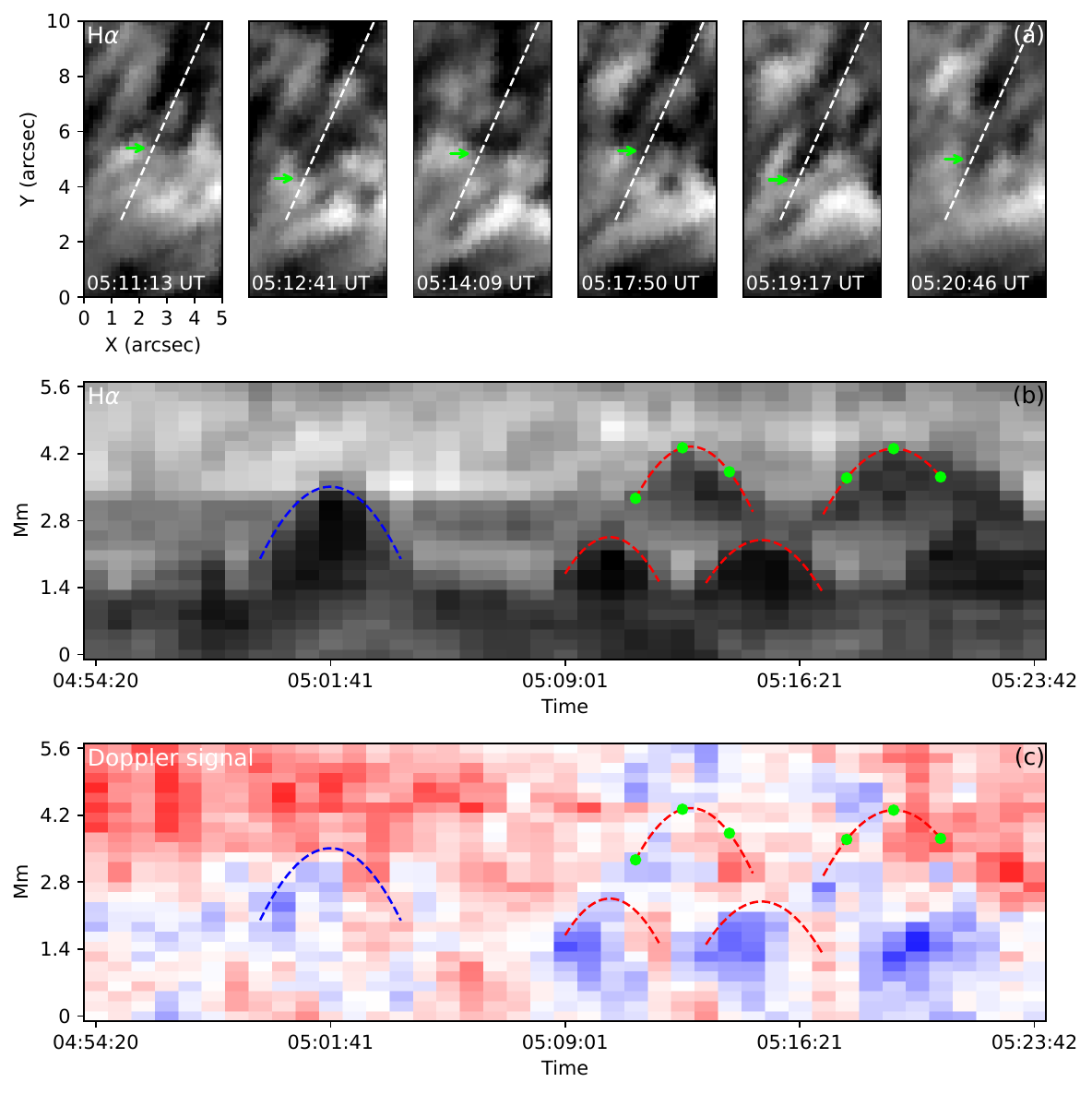}
    \caption{Examples of the evolution of DFs. (a) Time sequence of H$\alpha$ images corresponding to the white box~1 in Figure~\ref{fig:figure1}(b). The white dashed line indicates the position of the slice, and the green arrows mark the locations of the DFs. (b) Time–distance diagram obtained along the white dashed line in panel~(a). (c) Corresponding Doppler signal time–distance diagram. The blue dashed curve denotes the trajectory of a single-parabola event, while the red dashed curves denote the trajectories of two multiple-parabola events. The green dots correspond sequentially to the green arrows in panel~(a).}
    \label{fig:figure2}
\end{figure*}

\begin{figure*}[ht]
    \centering
    \includegraphics[width=0.64\linewidth]{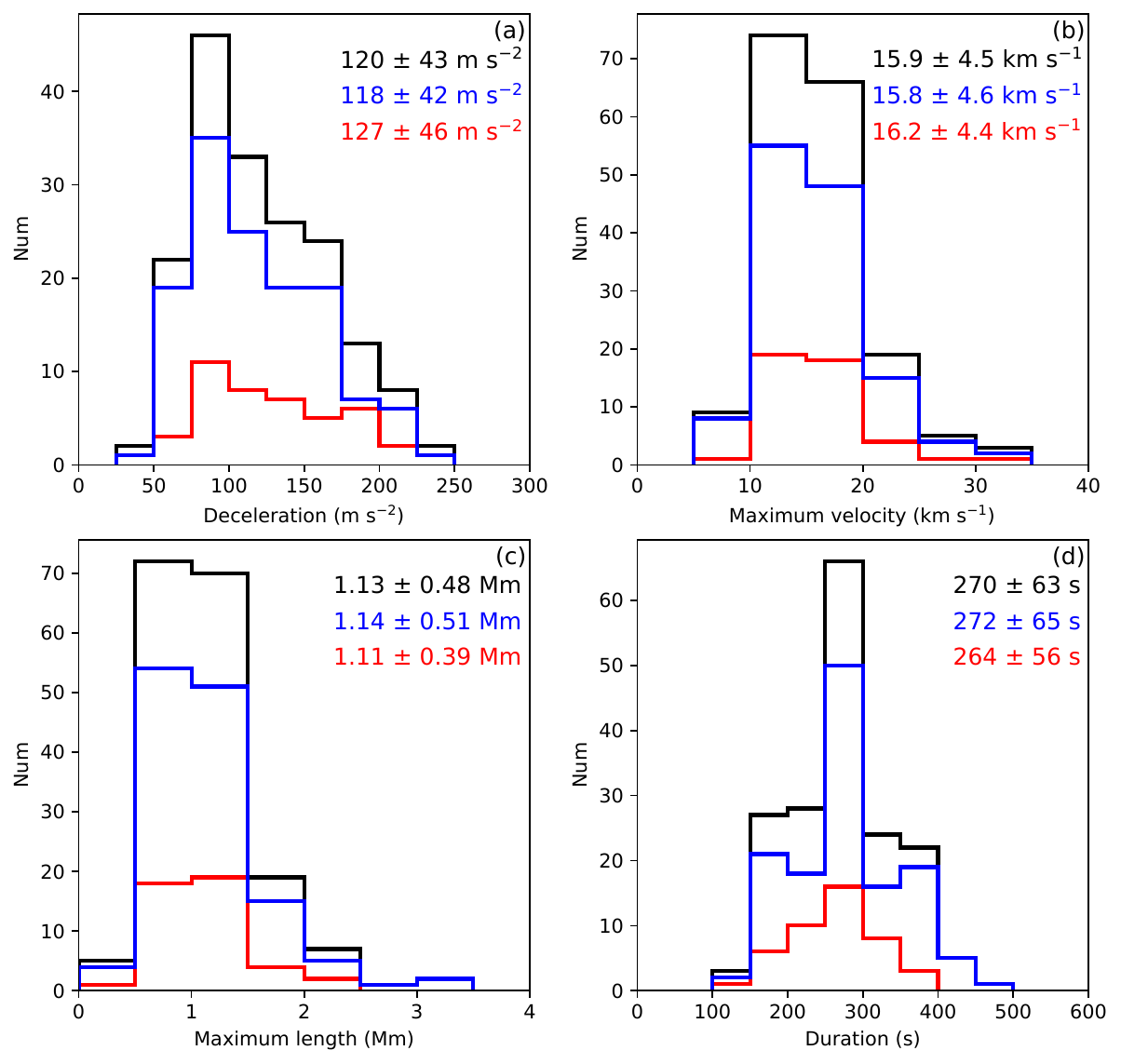}
    \caption{Histograms of the kinematic properties of DFs. (a) Distribution of decelerations. (b) Distribution of maximum velocities. (c) Distribution of maximum lengths. (d) Distribution of durations. The black histograms represent 176 DFs, while the blue and red histograms represent 132 single-parabola and 44 multiple-parabola events, respectively. The mean values and corresponding
    $1\sigma$ standard deviations for each group are annotated in the corresponding colors.}
    \label{fig:figure3}
\end{figure*}

\begin{figure*}[ht]
    \centering
    \includegraphics[width=0.72\linewidth]{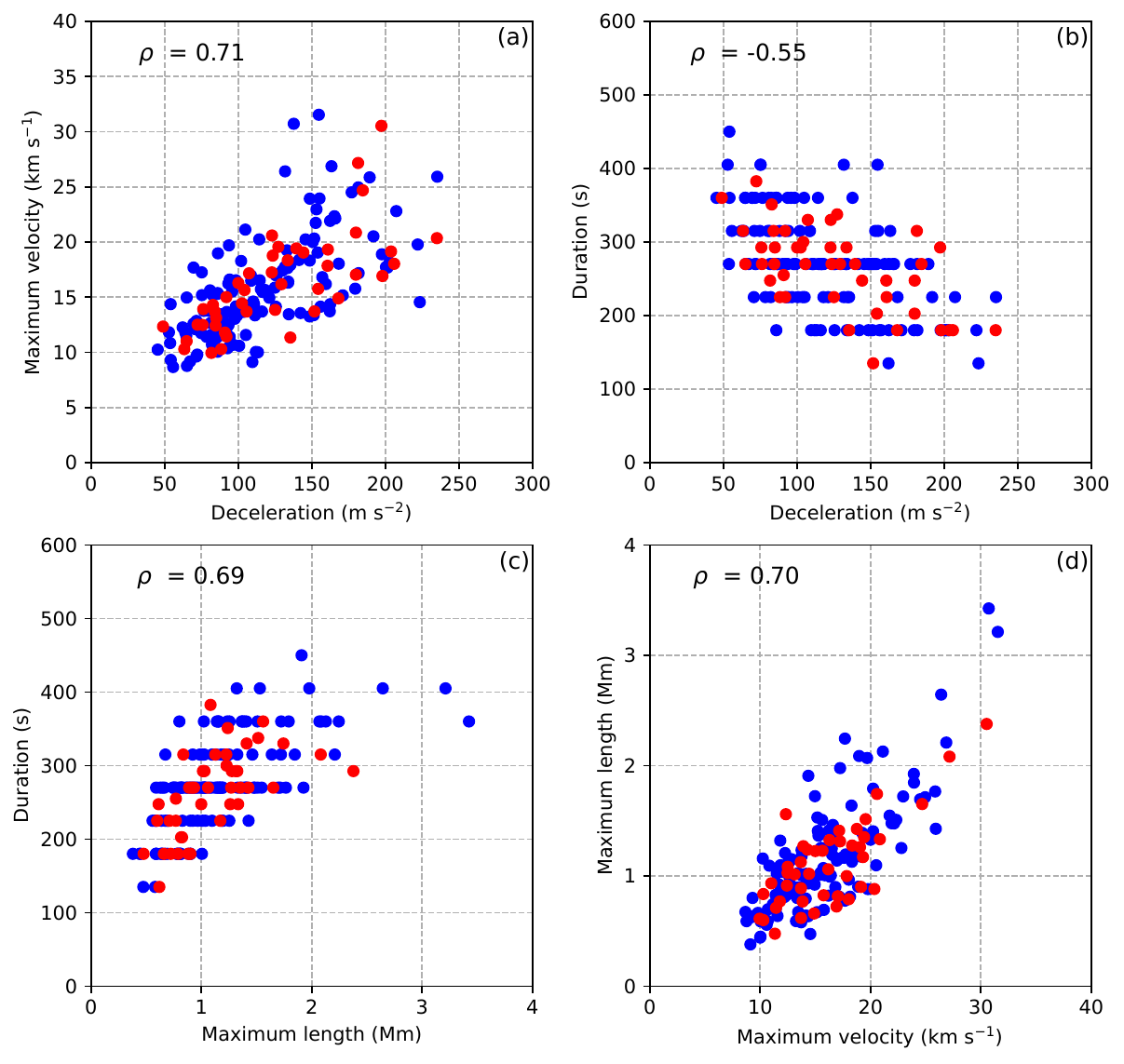}
    \caption{Scatter plots of the kinematic properties of DFs. (a) Deceleration versus maximum velocity. (b) Deceleration versus duration. (c) Maximum length versus duration. (d) Maximum velocity versus maximum length. Blue points denote single-parabola events, while red points denote multiple-parabola events. The Spearman correlation coefficients are annotated in each panel.}
    \label{fig:figure4}
\end{figure*}

\begin{figure*}[ht]
    \centering
    \includegraphics[width=0.80\linewidth]{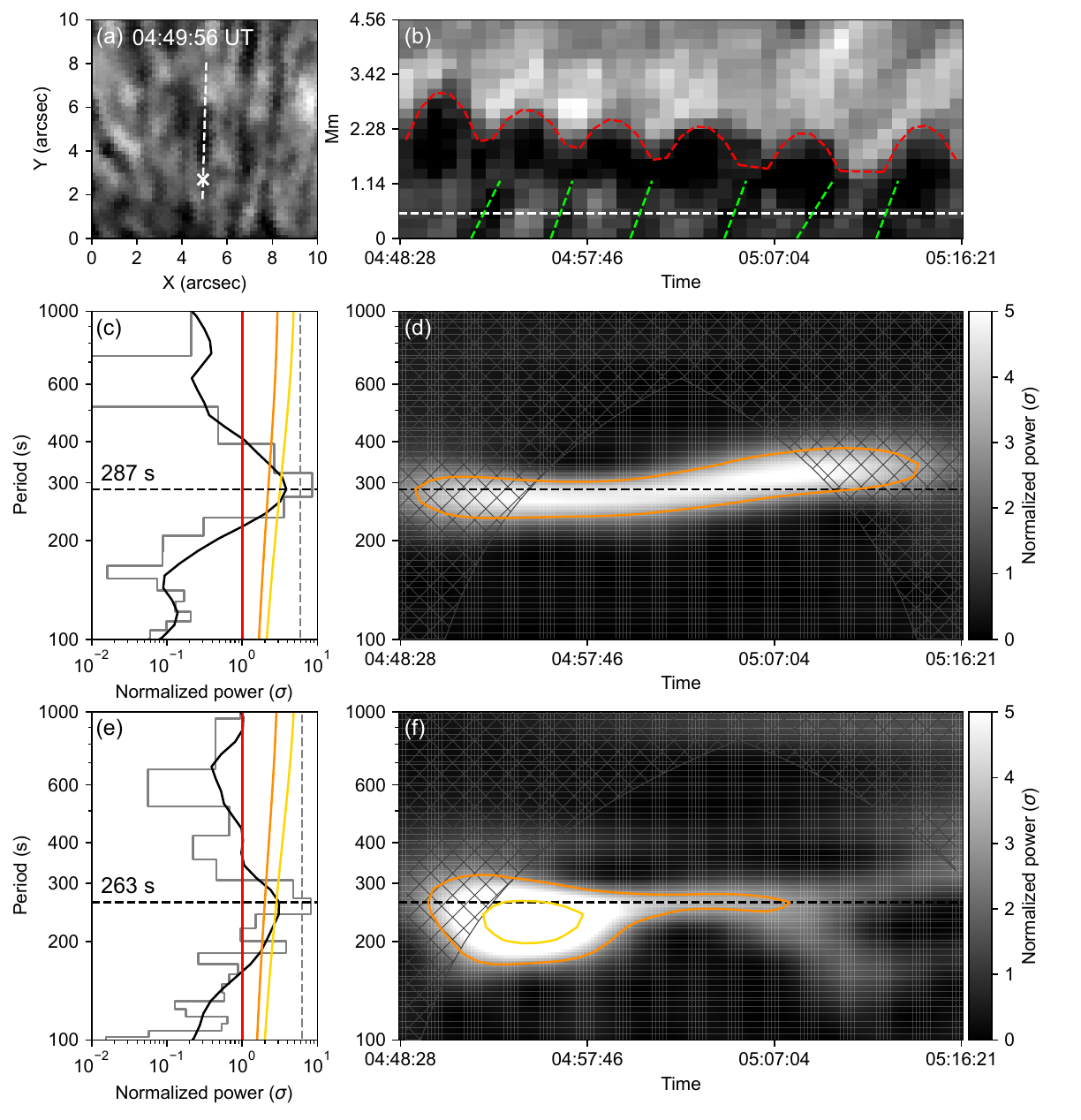}
    \caption{Periodicities of DFs and running penumbral waves. (a) H$\alpha$ image corresponding to the white box~2 in Figure~\ref{fig:figure1}(b). The white dashed line indicates the position of the slice, and the white cross marks the location of the white dashed line in panel~(b). (b) Time-distance diagram obtained along the white dashed line in panel~(a). The red dashed curve denotes the trajectory of the multiple-parabola event, while the green dashed lines denote the trajectories of the running penumbral waves. (c)–(d) Normalized time-averaged and wavelet power spectra of the multiple-parabola event. (e)–(f) Normalized time-averaged and wavelet power spectra of the running penumbral waves. The normalized time-averaged power spectra (black) are shown with the noise power (red) and the 95\% confidence levels (local: orange; global: yellow). The Fourier power spectra (gray) are shown with their 95\% confidence levels (gray dashed). The dominant oscillation periods are indicated by black dashed lines in panels (c)–(f). The hatched regions in the wavelet power spectra mark the cone of influence (COI).}
    \label{fig:figure5}
\end{figure*}

\begin{figure*}[ht]
    \centering
    \includegraphics[width=0.83\linewidth]{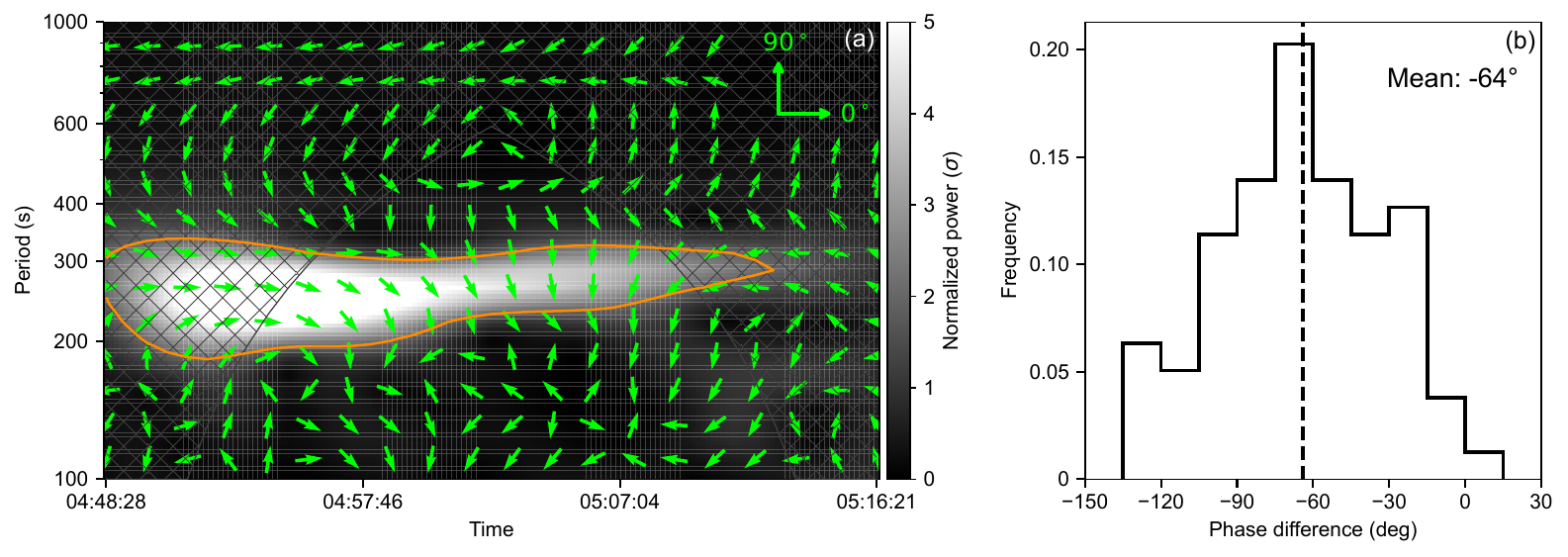}
    \caption{Phase difference of the multiple-parabola event relative to the running penumbral waves shown in Figure~\ref{fig:figure5}. (a) Wavelet cross-power spectrum of the two time series. The orange contour indicates the global 95\% confidence level, and the hatched regions mark the cone of influence (COI). The green arrows denote the local phase difference, with rightward and upward arrows corresponding to $0^\circ$ and $90^\circ$, respectively. (b) Normalized distribution of phase differences derived from regions above the global 95\% confidence level and outside the COI. The mean value is indicated by the black dashed line and annotated in the panel.}
    \label{fig:figure6}
\end{figure*}

\begin{figure*}[ht]
    \centering
    \includegraphics[width=0.79\linewidth]{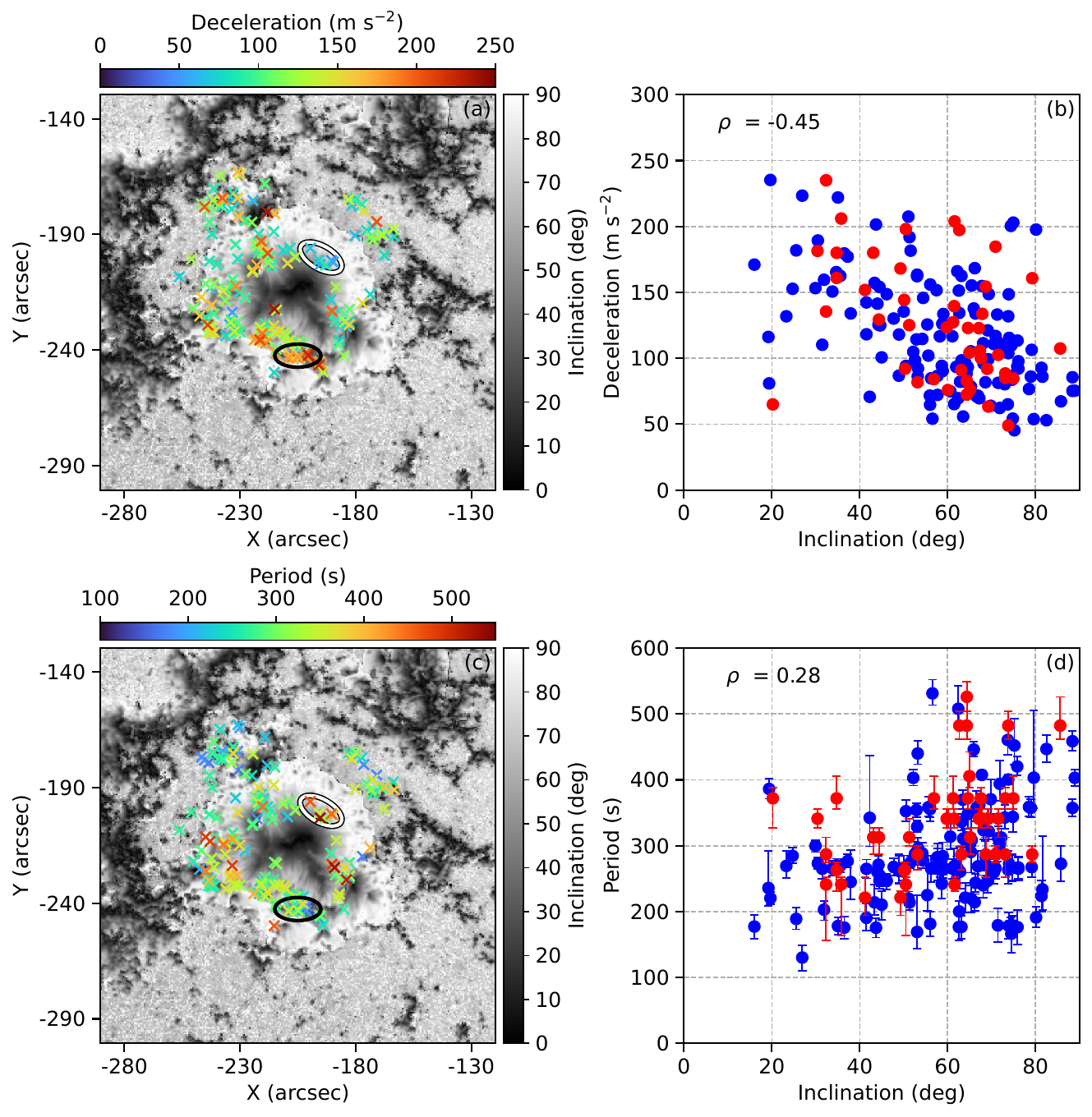}
    \caption{Variations of deceleration and period with the magnetic field inclination. (a) Spatial distribution of DFs overlaid on the inclination map, with colors indicating deceleration. The grayscale color bar indicates the inclination. (b) Inclination versus deceleration. (c) Spatial distribution of DFs overlaid on the inclination map, with colors indicating period. The grayscale color bar indicates the inclination. (d) Inclination versus period. The vertical error bars indicate the uncertainties in the periods. Black and white ellipses in panels~(a) and (c) mark representative regions. The Spearman correlation coefficients are annotated in panels~(b) and (d).}
    \label{fig:figure7}
\end{figure*}

\begin{figure*}[ht]
    \centering
    \includegraphics[width=\linewidth]{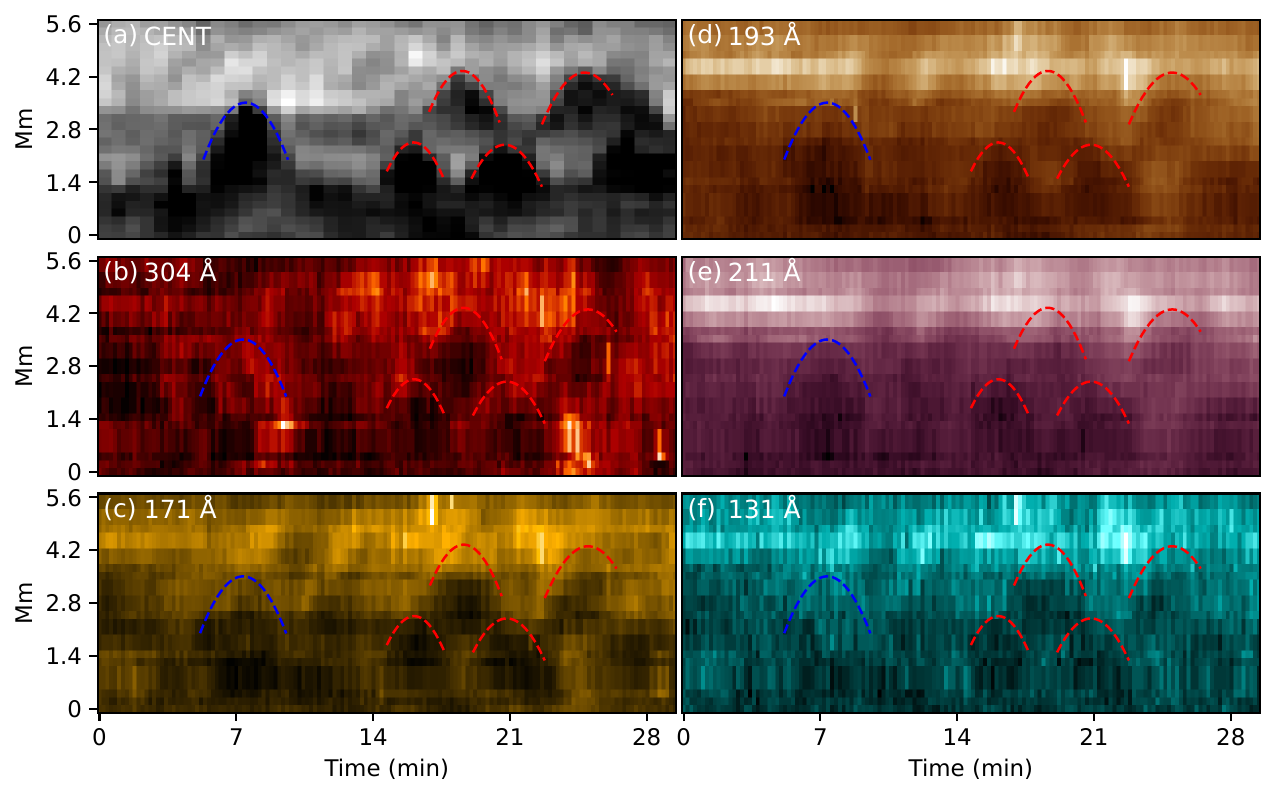}
    \caption{Time-distance diagrams showing the multiwavelength signatures of DFs. (a) Time-distance diagram in H$\alpha$ obtained along the white dashed line in Figure~\ref{fig:figure2}(a), which is the same as that shown in Figure~\ref{fig:figure2}(b). (b)--(f) Corresponding time-distance diagrams in 304, 171, 193, 211, and 131~\AA. The blue dashed curve denotes the trajectory of a single-parabola event, while the red dashed curves denote the trajectories of two multiple-parabola events.}
    \label{fig:figure8}
\end{figure*}

A widely accepted explanation for DFs is that they are driven by magnetoacoustic shocks \citep{2006ApJ...647L..73H}. One-dimensional MHD simulations by \citet{2004Natur.430..536D} show that photospheric oscillations leak into the chromosphere along inclined magnetic flux tubes, where they steepen into shocks that drive jet-like motions. \citet{2006ApJ...647L..73H} and \citet{2007ApJ...655..624D} compared the properties of DFs derived from observations and numerical simulations, finding excellent agreement that supports the shock-driven mechanism. In this scenario, chromospheric plasma experiences an impulsive acceleration followed by a nearly constant deceleration as the shock propagates, naturally producing the observed parabolic motions of DFs.

Understanding the origin of these shocks lies in the generation and propagation of slow magnetoacoustic waves in the solar atmosphere. Numerical simulations show that convective motions in and around magnetic flux concentrations can excite slow magnetoacoustic waves \citep{2011ApJ...730L..24K,2016ApJ...827....7K}. As slow magnetoacoustic waves propagate upward in the stratified atmosphere, their amplitudes rapidly increase with decreasing density, eventually steepening into shocks \citep{2006ApJ...640.1153C}. The propagation of slow magnetoacoustic waves is constrained by the acoustic cutoff frequency, which varies with height due to changes in the local plasma conditions \citep{2015A&A...582A..57A,2018A&A...617A..39F}. In addition, advanced numerical simulations show that long-period waves preferentially propagate in regions with strong, inclined magnetic fields, whereas short-period waves dominate in regions with weaker or more vertical magnetic fields \citep{2007ApJ...666.1277H,2011ApJ...743..142H}. This result provides a natural explanation for the observed regional differences in the properties of DFs, which are driven by waves with different periods \citep{2007ApJ...655..624D,2013ApJ...776...56R}.

Considerable efforts have been devoted to investigating the multiwavelength signatures of DFs, revealing their responses across different layers of the solar atmosphere. \citet{2008ApJ...673.1194L} presented cospatial and cotemporal chromospheric observations in H$\alpha$ and Ca~II wavebands, showing distinctive Doppler shifts of DFs. Subsequent studies have identified enhanced emission associated with DFs in the transition region, indicating the impact and heating of chromospheric shocks \citep{2016ApJ...817..124S}. More recently, high-resolution observations from Solar Orbiter have revealed coronal counterparts of DFs as small, bright blob-like features exhibiting parabolic motions, which are interpreted as the hot tips of DFs \citep{2023A&A...670L...3M,2023A&A...678L...5M}.

The mechanism by which slow magnetoacoustic waves steepen into shocks to drive DFs has been well established \citep{2006ApJ...647L..73H,2007ApJ...655..624D}. However, direct observational evidence remains relatively limited. In addition, the responses of DFs over an extended range of atmospheric heights remain to be further clarified.

In this paper, we present a statistical analysis of 176 DFs to investigate their kinematic properties and driving mechanism. Section~\ref{sec:section2} describes the observations and data reduction methods. Section~\ref{sec:section3} provides the detailed results of our analysis. Section~\ref{sec:section4} summarizes our findings and discusses the remaining issues.

\section{Observations and Data Reduction} \label{sec:section2}

\subsection{Observations} \label{subsec:section2.1}

Figure~\ref{fig:figure1}(a) shows the full-disk H$\alpha$ image from the Chinese H$\alpha$ Solar Explorer (CHASE, \citet{2022SCPMA..6589602L,2022SCPMA..6589603Q}). The CHASE satellite is equipped with the H$\alpha$ Imaging Spectrograph (HIS) for observations in the H$\alpha$ and Fe~I wavebands in raster scanning mode (RSM). Active region NOAA 13182 is located at S15E11. It is dominated by a mature negative-polarity sunspot with a well-defined umbra and penumbra. High-resolution chromospheric observations in the H$\alpha$ waveband are provided by the New Vacuum Solar Telescope (NVST, \citet{2014RAA....14..705L,2020ScChE..63.1656Y}), as shown in Figure~\ref{fig:figure1}(b). An adaptive optics system is used to reduce the effects of atmospheric seeing and improve the image quality \citep{2016ApJ...833..210R,2023SCPMA..6669611Z}. The observations analyzed in this study cover the period from 04:39:40 to 05:50:52 UT on 2023 January 10. This interval is characterized by relatively good seeing and effective adaptive optics correction, yielding high-resolution H$\alpha$ images centered on the sunspot. The field of view of the H$\alpha$ image is $150^{\prime\prime} \times 150^{\prime\prime}$, with a spatial resolution of 0\farcs165 pixel$^{-1}$ and a cadence of 45~s. In this work, we manually identify 176 DFs from the NVST time-series observations. Figure~\ref{fig:figure1}(c) shows the corresponding 304\,\AA\ image from the Solar Dynamics Observatory (SDO, \citet{2012SoPh..275....3P}). The SDO provides seven extreme-ultraviolet wavebands from the Atmospheric Imaging Assembly (AIA, \citet{2012SoPh..275...17L}) and vector magnetograms from the Helioseismic and Magnetic Imager (HMI, \citet{2012SoPh..275..207S}).

\subsection{Data Reduction Methods} \label{subsec:section2.2}

We identify DFs based on their clear up-and-down motions in the H$\alpha$ images and their distinct parabolic trajectories in the corresponding time--distance diagrams. A total of 176 DFs are identified from the NVST observations, comprising 132 single-parabola events and 44 multiple-parabola events. Multiple-parabola events are defined as those consisting of at least two consecutive parabolic motions. The interval between successive motions should not exceed two frames; otherwise, they are classified as single-parabola events. As shown in Figure~\ref{fig:figure1}(b), single-parabola events are denoted by blue crosses, while multiple-parabola events are denoted by red crosses.

The trajectories of DFs are fitted with parabolic functions to derive their kinematic properties: deceleration, maximum velocity, maximum length, and duration. The deceleration is given by the absolute value of the constant acceleration obtained from the parabolic fit. The maximum velocity corresponds to the initial velocity at the starting point. The maximum length is measured as the height difference between the vertex and the starting point. The duration is calculated as the time interval between the first and last frames in which the DFs are visible. It should be noted that the measured parameters are projected components on the plane of the sky. The parameters of multiple-parabola events are determined by averaging those of the individual parabolic motions. The correlations between the parameters are evaluated using the Spearman rank correlation coefficient.

For single-parabola events, despite the absence of explicit periodicity, the period $P$ is estimated from $P = 2v_{\max}/a$, where $v_{\max}$ is the maximum velocity and $a$ is the deceleration. This analytical relation is derived from the magnetoacoustic shock model \citep{2014CEAB...38...39K}. For multiple-parabola events, the period is determined from wavelet analysis of the trajectories of DFs \citep{1998BAMS...79...61T}. Following \citet{2016ApJ...825..110A}, the background noise model includes kappa, power-law, and constant components; here, the kappa term is set to zero, which remains adequate for capturing the dominant oscillation period. The method has also been adopted in previous studies \citep{2022A&A...664A..28M,2026A&A...706A...1D}. The uncertainties in the derived periods are estimated separately for single-parabola and multiple-parabola events. For single-parabola events, we add Gaussian random errors with a standard deviation of one pixel to the measured trajectory positions. The trajectory is then fitted again with a parabola to calculate a new period. This procedure is repeated 10,000 times for each event, and the period uncertainty is determined from the 16th–84th percentile range of the resulting periods. For multiple-parabola events, the periods are derived from wavelet analysis, with uncertainties from measurement errors, the finite observation window, and the discrete wavelet period grid. Accordingly, we perform injection–recovery tests for three input periods: the derived period and its lower and upper half-grid boundaries. For each input period, the trajectory is rescaled to match that period, and Gaussian random errors with a standard deviation of one pixel are added to the trajectory positions. The period is then recovered using the same wavelet analysis. This procedure is repeated 1,000 times for each input period. The period uncertainty is determined from the envelope of the 16th–84th percentile ranges obtained for the three input periods. The wavelet cross-power analysis is performed to examine the phase difference between the trajectory time series of a multiple-parabola event and the intensity time series of the associated running penumbral waves, following the method described by \citet{2004NPGeo..11..561G}.

To characterize the Doppler signal of DFs, the intensity difference between the blue ($-0.6\,\mathrm{\mathring{A}}$) and red ($+0.6\,\mathrm{\mathring{A}}$) wings of the H$\alpha$ line is used as a proxy for the line-of-sight velocity \citep{2000NewA....5....1T}. To investigate the magnetic field inclination of DFs, the vector magnetic field data from HMI are used to calculate the angle relative to the local normal direction, ranging from $0^\circ$ for vertical fields to $90^\circ$ for horizontal fields.

\section{Results} \label{sec:section3}

As shown in Figure~\ref{fig:figure1}(b), the 176 DFs are distributed from the penumbra to the superpenumbra of the sunspot. The up-and-down motions of DFs are aligned with the penumbral fibrils, which indicate the direction of the magnetic field. Such motions sometimes occur consecutively, exhibiting their periodicity. A time sequence of H$\alpha$ images is shown in Figure~\ref{fig:figure2}(a), where the green arrows mark the locations of the DFs. The multiple-parabola event consists of two consecutive up-and-down motions, corresponding to the two parabolic trajectories denoted by the red dashed curves with green dots in Figure~\ref{fig:figure2}(b). The time–distance diagram shows a single-parabola event and two multiple-parabola events, denoted by the blue and red dashed curves, respectively. In Figure~\ref{fig:figure2}(c), the Doppler signal time–distance diagram shows that DFs exhibit blueshifts during the upward phase and redshifts during the downward phase. This behavior is consistent with the Ca II observations of \citet{2013ApJ...776...56R}, in which different line-wing positions were used to trace different phases in the evolution of DFs.

\subsection{Kinematic Properties} \label{subsec:section3.1}

Based on the definitions and methods described in Section~\ref{subsec:section2.2}, the derived kinematic properties of the 176 DFs are shown in Figure~\ref{fig:figure3}. The black histograms represent 176 DFs, while the blue and red histograms represent 132 single-parabola and 44 multiple-parabola events, respectively.  Figure~\ref{fig:figure3}(a) shows the distribution of deceleration, which ranges from 45 to 235~m~s$^{-2}$ with a mean value of $120 \pm 43$~m~s$^{-2}$. The mean values are $118 \pm 42$~m~s$^{-2}$ for single-parabola events and $127 \pm 46$~m~s$^{-2}$ for multiple-parabola events. The results indicate that the deceleration of DFs is lower than the solar gravitational acceleration of 274~m~s$^{-2}$. After accounting for projection effects, this behavior can be naturally understood in terms of plasma motions along inclined magnetic fields, where the effective gravity is the component of solar gravity projected along the field direction. Figure~\ref{fig:figure3}(b) shows the distribution of maximum velocity, which ranges from 8.7 to 31.5~km~s$^{-1}$ with a mean value of $15.9 \pm 4.5$~km~s$^{-1}$. The mean values are $15.8 \pm 4.6$~km~s$^{-1}$ for single-parabola events and $16.2 \pm 4.4$~km~s$^{-1}$ for multiple-parabola events. The results indicate that the initial velocity of DFs exceeds the local sound speed in the chromosphere, allowing for shock formation. The maximum length of DFs ranges from 0.38 to 3.42~Mm, with an average of $1.13 \pm 0.48$~Mm, as shown in Figure~\ref{fig:figure3}(c). The mean maximum lengths are $1.14 \pm 0.51$~Mm and $1.11 \pm 0.39$~Mm for single-parabola and multiple-parabola events, respectively. The mean values are comparable to the typical lengths of DFs and spicules \citep{2007ApJ...655..624D,2012ApJ...759...18P}. The duration of DFs ranges from 135 to 450~s, with an average of $270 \pm 63$~s, as shown in Figure~\ref{fig:figure3}(d). The mean durations are $272 \pm 65$~s and $264 \pm 56$~s for single-parabola and multiple-parabola events, respectively. The mean values are comparable to the 5-minute period that dominates chromospheric penumbral oscillations. Taken together, these distributions suggest that single-parabola and multiple-parabola events exhibit consistent kinematic properties.

Some of the kinematic properties are clearly correlated with each other. The scatter plot in Figure~\ref{fig:figure4}(a) shows a strong positive correlation between deceleration and maximum velocity, with a Spearman correlation coefficient of 0.71. The positive correlation indicates that DFs with larger decelerations tend to have higher maximum velocities. According to previous numerical simulations \citep{2007ApJ...666.1277H}, this significant correlation arises from shock wave–driven chromospheric jets. In contrast, Figure~\ref{fig:figure4}(b) shows that DFs with larger decelerations tend to have shorter durations ($\rho = -0.55$). In Figure~\ref{fig:figure4}(c), DFs with greater maximum lengths are generally associated with longer durations ($\rho = 0.69$). A similar trend is seen in Figure~\ref{fig:figure4}(d), where higher maximum velocities correspond to greater maximum lengths ($\rho = 0.70$). The correlations are generally moderate to strong, and single-parabola and multiple-parabola events exhibit uniform distributions.

\subsection{Driving Mechanism} \label{subsec:section3.2}

Figure~\ref{fig:figure5} shows the wavelet analysis of a multiple-parabola event and running penumbral waves. The time-distance diagram shown in Figure~\ref{fig:figure5}(b) is obtained along the white dashed line in panel (a). The multiple-parabola event with six parabolic motions is denoted by the red dashed curve. This event is located close to the umbra–penumbra boundary. The running penumbral waves are denoted by the green dashed lines. A clear spatiotemporal relationship is found between the multiple-parabola event and the running penumbral waves. The waves propagate outward from the umbra–penumbra boundary and reach the locations corresponding to the starting points of the parabolic trajectories. Since running penumbral waves are generally regarded as signatures of slow magnetoacoustic waves, this relationship suggests a physical connection between DFs and slow magnetoacoustic waves.

Wavelet analysis is performed to examine the periodicities of the multiple-parabola event and the running penumbral waves. The normalized time-averaged power spectrum of the multiple-parabola event and the corresponding wavelet power spectrum are shown in Figure~\ref{fig:figure5}(c)–(d). The normalized power spectrum reveals a reliable dominant oscillation period of 287~s. In the wavelet power spectrum, a significant wave packet above the global 95\% confidence level persists throughout the entire time interval. The intensity time series extracted along the white dashed line in Figure~\ref{fig:figure5}(b) is analyzed to determine the periodicity of the running penumbral waves. The normalized time-averaged power spectrum and the corresponding wavelet power spectrum are shown in Figure~\ref{fig:figure5}(e)-(f). The results indicate a dominant oscillation period of 263~s for the running penumbral waves, accompanied by a corresponding wave packet in the wavelet power spectrum. The Fourier power spectra in panels~(c) and (e) also show peak powers at the dominant oscillation periods, further confirming the reliability of the detected periodicities. The agreement between the dominant oscillation period of the multiple-parabola event and that of the running penumbral waves supports the interpretation that DFs are driven by slow magnetoacoustic waves.

The wavelet cross-power analysis is a useful tool for characterizing the phase difference between two time series that are expected to be related. Figure~\ref{fig:figure6}(a) shows the wavelet cross-power spectrum of the two time series shown in Figure~\ref{fig:figure5}. The green arrows denote the local phase difference of the multiple-parabola event relative to the running penumbral waves. In this spectrum, the regions above the global 95\% confidence level and outside the cone of influence (COI) exhibit strong coherent power, where the phase difference is considered reliable. These regions are dominated by downward arrows, indicating that the multiple-parabola event lags behind the running penumbral waves. This behavior is a natural consequence of the propagation of slow magnetoacoustic waves that drive the DFs. As shown in Figure~\ref{fig:figure6}(b), the phase differences range from $-135^\circ$ to $5^\circ$, with a mean value of $-64^\circ$. The propagation time is estimated as $\tau=\phi P/2\pi$, where $\phi$ is the phase difference and $P$ is the oscillation period. Assuming a illustrative distance of $d=1~\mathrm{Mm}$, substituting $\phi=64^\circ$ and $P=275~\mathrm{s}$ gives an estimated phase speed of $v=d/\tau \approx 20~\mathrm{km~s^{-1}}$. The value is sufficient for shock formation and is comparable to the maximum velocities of DFs; however, it should be emphasized that it is an estimate rather than the true propagation speed of waves.

The vector magnetic field data from HMI are used to derive the magnetic field inclination. To further constrain the behavior of DFs, the variations of deceleration and period with magnetic field inclination are investigated. The results are presented in Figure~\ref{fig:figure7}. In Figures~\ref{fig:figure7}(a) and (c), the colors of the crosses indicate the deceleration and period of DFs, respectively. The periods of single-parabola events are derived from the analytical relation described in Section~\ref{subsec:section2.2}, while those of multiple-parabola events are obtained from wavelet analysis. The black and white ellipses mark representative regions with contrasting magnetic field inclinations. The black ellipses correspond to regions with relatively vertical magnetic fields, where DFs exhibit larger decelerations and shorter periods. In contrast, the white ellipses correspond to regions with relatively horizontal magnetic fields, where DFs exhibit smaller decelerations and longer periods. As shown in the scatter plots in Figures~\ref{fig:figure7}(b) and (d), deceleration tends to decrease with increasing inclination ($\rho = -0.45$), while the period tends to increase with inclination ($\rho = 0.28$). Although the correlations are weak to moderate, these regional differences are consistent with the scenario in which DFs are driven by slow magnetoacoustic waves with different periods propagating along magnetic fields, in agreement with previous observations and numerical simulations \citep{2007ApJ...655..624D,2011ApJ...743..142H}.

\subsection{Multiwavelength Signatures} \label{subsec:section3.3}

Figure~\ref{fig:figure8} shows the multiwavelength signatures of DFs, revealing their responses over an extended range of atmospheric heights. As indicated by the dashed curves, corresponding responses of the DFs are observed from the chromosphere to the transition region and corona. The time–distance diagram in H$\alpha$ shows the DFs as dark features, and associated signatures can be traced in the 304, 171, 193, 211, and 131~\AA\ wavebands. Recent high-resolution observations have revealed that DFs can appear as blob-like hot tips and bright parabolic trajectories in transition-region and coronal diagnostics \citep{2023A&A...670L...3M,2023A&A...678L...5M}. Similar features can be identified in the 171~\AA\ waveband, although they are weak or ambiguous because of limited spatial resolution and complex background emission.

In light of the analysis in Section~\ref{subsec:section3.2}, slow magnetoacoustic waves originating from the photosphere propagate upward and steepen into shocks in the chromosphere. These shocks can drive DFs and may also transport energy to higher atmospheric layers, where it could be dissipated. Earlier numerical studies have shown that chromospheric shocks can propagate upward and lift the transition region \citep{1982SoPh...75...99S,1982SoPh...77..121S,1982ApJ...257..345H,1988ApJ...327..950S,1990ApJ...349..647S,1994SSRv...70..103H,1994A&A...290..995H}. More recently, the numerical simulations of \citet{2016ApJ...827....7K} further suggest that such shocks can drive DFs and generate compressive disturbances that extend into the corona. Thus, the multiwavelength signatures of DFs from the chromosphere to the corona imply a potential contribution to energy transport and heating in the solar atmosphere.

\section{Conclusions and Discussion} \label{sec:section4}

In this study, we perform a statistical analysis of 176 DFs identified from high-resolution NVST H$\alpha$ observations. Based on the continuity of their up-and-down motions, these DFs are classified into 132 single-parabola events and 44 multiple-parabola events. The kinematic properties indicate that the DFs typically have a deceleration of about $120~\mathrm{m~s^{-2}}$, a maximum velocity of about $15.9~\mathrm{km~s^{-1}}$, a maximum length of about $1.13~\mathrm{Mm}$, and a duration of about $270~\mathrm{s}$. Several parameters are moderately to strongly correlated. The most prominent relation is the positive correlation between deceleration and maximum velocity, with larger decelerations generally corresponding to higher maximum velocities, consistent with the shock-driven scenario \citep{2014ApJ...786..137T,2017ApJ...838....2Z}. The multiple-parabola event shows a clear spatiotemporal association with the running penumbral waves, which propagate outward from the umbra–penumbra boundary toward the starting points of the parabolic trajectories. The dominant oscillation period of the event agrees well with that of the running penumbral waves, as revealed by wavelet analysis. Similar links between penumbral waves and chromospheric dynamic features have also been reported for superpenumbral fibrils and fan-shaped chromospheric jets \citep{2014ApJ...789..108C,2018ApJ...855L..19R}. Running penumbral waves are generally interpreted as the chromospheric signatures of slow magnetoacoustic waves propagating upward along inclined magnetic field lines \citep{2007ApJ...671.1005B,2013ApJ...779..168J}. The origin of these waves is commonly attributed to the upward leakage of photospheric p-mode oscillations into the chromosphere \citep{2004Natur.430..536D,2015ApJ...800..129M}. Alternatively, \citet{1982ApJ...257..345H} proposed that a single photospheric disturbance can generate a series of rebound shocks that repeatedly propagate upward through the chromosphere. These results support the interpretation that DFs are driven by slow magnetoacoustic waves. The magnetic field inclination analysis reveals distinct regional differences in the behavior of DFs. Larger decelerations and shorter periods are found in regions with more vertical magnetic fields, whereas smaller decelerations and longer periods occur in regions with more horizontal magnetic fields. These regional differences suggest that the behavior of DFs is influenced by the acoustic cutoff frequency. This frequency regulates the propagation of waves with different periods along magnetic fields with different inclinations. The multiwavelength signatures of DFs are detected from the chromosphere to the transition region and corona, indicating that DFs can be traced in higher atmospheric layers. These responses point to a potential contribution of DFs to energy transport and heating in the solar atmosphere.

These results provide direct observational evidence that reinforces the explanation that DFs are driven by slow magnetoacoustic waves and indicates that their responses can be detected over a broad range of atmospheric heights. However, a number of remaining issues need to be clarified.

Among these issues, the limited temporal and spatial resolution of the observations may prevent smaller-scale DFs from being fully resolved. In sunspot umbrae, unresolved or only partially resolved small-scale DFs may also exist, as high-resolution observations have revealed dynamic fine structures with subarcsecond widths and lifetimes ranging from several tens of seconds to a few minutes \citep{2014ApJ...787...58Y,2017A&A...605A..14N,2020A&A...642A.215H}. Another uncertainty concerns the determination of the periods of DFs. For single-parabola events, the periods are inferred from the analytical relation based on idealized shock-driven motion, which may not fully capture the complexity of the real solar atmosphere. For multiple-parabola events, most samples contain only two or three consecutive parabolic trajectories, which limits the temporal coverage of the wavelet analysis. Therefore, the derived periods should be regarded as approximate values. In addition, the magnetic field inclination is calculated using the photospheric vector magnetic field data from HMI, whereas a more accurate determination of the field inclination relevant to DFs requires precise chromospheric magnetic field measurements. This limitation may partly account for the weak to moderate correlations shown in Figure~\ref{fig:figure7}.

Beyond these issues, detailed observations of the relationship between DFs and shocks are needed to establish a more complete observational picture of how slow magnetoacoustic waves steepen into shocks that drive DFs \citep{2008ApJ...673.1194L,2016ApJ...817..124S}. Moreover, Alfvénic waves carried by DFs deserve further investigation, as they may provide an important channel for energy transport and heating in the solar atmosphere \citep{2007Sci...318.1574D,2017Sci...356.1269M}. In this context, previous numerical studies have suggested that Alfvén waves may produce magnetoacoustic disturbances or shocks that affect the dynamics of the transition region \citep{1982SoPh...75...35H,1985ApJ...296..746M,1999ApJ...514..493K}. Addressing these questions requires comprehensive multiwavelength observations with improved spatial and temporal resolution.

\begin{acknowledgments}

The authors are grateful to the CHASE, NVST, and SDO teams for providing the data. This work is supported by the Strategic Priority Research Program of the Chinese Academy of Sciences, grant No. XDB0560000; the National Key R\&D Program of China, under No. 2024YFA1612001; the National Science Foundation of China (NSFC) under Nos. 12325303, 12473059, 12433010, and 12403065; Yunnan Key Laboratory of Solar Physics and Space Science under No. 202205AG070009; the Yunnan Science Foundation of China under Nos. 202501AW070002 and 202601AS070076; the Yunnan Province Xing Dian Talent Support Program, the Yunling Scholar Project, and Yunnan Revitalization Talent Support Program under Nos. 202305AS350029, 202305AT350005.

\end{acknowledgments}

\clearpage

\bibliography{References}{}
\bibliographystyle{aasjournalv7}

\end{document}